\documentclass[reprint,footinbib,amsmath,amssymb,aps,pre,%
  floatfix,superscriptaddress,a4paper,nolongbibliography]{revtex4-2}
\usepackage{newtxtext,newtxmath}
\usepackage{microtype}
\usepackage{gensymb}
\usepackage[colorlinks=true,allcolors = blue]{hyperref}
\usepackage{color}
\usepackage{graphicx}

\newcommand{\dd}{\mathrm{d}}

\newcommand{\smallfrac}[2]{{\textstyle\frac{#1}{#2}}}

\newcommand{\order}[1]{\mathcal{O}{(#1)}}
\newcommand{\Outer}{{\mathrm{O}}}
\newcommand{\Inner}{{\mathrm{I}}}

\newcommand{\xc}{x_c}
\newcommand{\zc}{z_c}

\newcommand{\Eq}[1]{Eq.~\eqref{#1}}
\newcommand{\Eqs}[1]{Eqs.~\eqref{#1}}
\newcommand{\Fig}[1]{Fig.~\ref{#1}}

\newcommand{\Refcite}[1]{Ref.~\onlinecite{#1}}

\newcommand{\Table}[1]{Table~\ref{#1}}

\newcommand{\naive}{na\"\i{}ve}

\newcommand{\latin}[1]{{\itshape #1}}

\newcommand{\via}{\latin{via}}

\newcommand{\eg}{e.g.}
\newcommand{\ie}{i.e.}
\newcommand{\cf}{cf.}

\begin{document}

\title{Renormalisation group theory applied to $\ddot{x}+\dot{x}+x^2=0$}

\author{Joshua F. Robinson}
\email{joshua.robinson@stfc.ac.uk}
\affiliation{The Hartree Centre, STFC Daresbury Laboratory, Warrington, WA4 4AD, United Kingdom}
\affiliation{H.\ H.\ Wills Physics Laboratory, University of Bristol, Bristol BS8 1TL, United Kingdom}
\affiliation{Institut f\"ur Physik, Johannes Gutenberg-Universit\"at Mainz, Staudingerweg 7-9, 55128 Mainz, Germany}

\author{Patrick B. Warren}
\email{patrick.warren@stfc.ac.uk}
\affiliation{The Hartree Centre, STFC Daresbury Laboratory, Warrington, WA4 4AD, United Kingdom}
\affiliation{SUPA and School of Physics and Astronomy, The University of Edinburgh, Peter Guthrie Tait Road, Edinburgh EH9 3FD, United Kingdom}

\date{\today}

\begin{abstract}
  The titular ordinary differential equation (ODE) is encountered in the theory of on-axis inertial particle capture by a blunt stationary collector at a viscous-flow stagnation point.  Phase space for the ODE divides into two attractor basins, representing particle trajectories which do or do not collide with the collector in a finite time.   Written as $\ddot{x} + \dot{x} + \epsilon x^2 = 0$, we formulate the renormalisation group (RG) amplitude equations for this problem and argue that the critical trajectory which separates the attractor basins corresponds to a trivial but exact solution of these, and can therefore be extracted as a power series in $\epsilon$.  We show how this can be used to find the cross-over between capture and non-capture as a function of distance from the stagnation point, for a particle released into the flow with no initial acceleration.  This cross-over, previously only computable by numerical integration of the ODE, can therefore be expressed as a (numerically) convergent series with rational terms.
\end{abstract}

\maketitle

\section{Introduction}
In the theory of on-axis inertial particle capture at a stagnation point in viscous flow around a blunt stationary collector, the flow velocity approaching the stagnation point at $x = 0$ from the right scales as $-x^2$, assuming incompressibility and no-slip boundary conditions \cite{pozrikidis_2011, robinson2024}.  Assuming Stokes drag, the limiting equation of motion for a point particle reads~\cite{robinson2021}
\begin{equation}
m \ddot x = - \gamma (\dot{x} + U_0 x^2)\,,
\end{equation}
at leading order.  The particle mass $m$, drag coefficient $\gamma$ and characteristic flow velocity $U_0$ can be reabsorbed into rescalings of position $x$ and time $t$, giving the titular equation:
\begin{equation}\label{eq:ode}
\ddot{x} + \dot{x} + x^2 = 0\,.
\end{equation}
This deceptively simple extension of the simple harmonic oscillator exhibits a rich dynamics which are exemplary of a whole class of systems involving particle capture at low Reynolds number. A non-exhaustive list includes filtration of aerosol droplets by face masks \cite{wang2013, *howard2020, robinson2021}, water extraction \cite{parker2001,*mitchell2020, *shahrokhian2020, *azeem2020}, pollination \cite{pawu1989, *niklas1985}, marine life feeding on plankton \cite{boudina2020, *espinosa-gayosso2021, *palmer2004} and drug delivery \cite{finlay2019, *darquenne2020}. 

\begin{figure}[b]
  \centering
  \includegraphics[width=\linewidth]{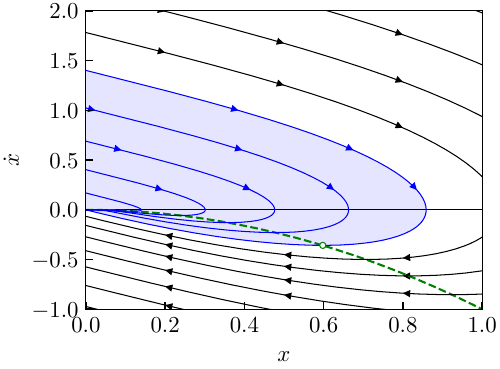}
  \caption{
    Particle trajectories following \Eq{eq:ode} for different initial conditions.
    Trajectories in the blue shaded region tend to $x \to 0$ as $t \to \infty$, whereas black trajectories pass through $x = 0$ and diverge to $x \to -\infty$ in finite time.
    The boundary of this blue region is the separatrix.
    The separatrix intersects the nullcline $\dot{x} = -x^2$ (green dashed line) at the point $x = \xc\simeq0.597$ (green circle).
  }
  \label{fig:portrait}
\end{figure}

To the best of our knowledge, \Eq{eq:ode} has no known analytic solution.  The phenomenology has been explored numerically in two recent works \cite{vallee2018, robinson2024}.
The system has a degenerate attractor at the origin in phase space $\ddot{x}=\dot{x}=0$ which is reached by a set of trajectories (blue region in \Fig{fig:portrait}) after an infinite time, but also exhibits trajectories which diverge to $x\to-\infty$.
The former represent particle trajectories which never collide with the collector (except at $t \to \infty$) whereas the latter represents trajectories which do collide in a finite time.
Phase space $(x, \dot{x})$ is foliated by these trajectories and divides into two attractor basins, corresponding to the two possible fates.
For instance, if a particle is injected into the flow with zero initial acceleration, it never reaches the stagnation point if $x(t=0) < \xc \simeq 0.597$, but collides with the surface in a finite time if $x(t=0) > \xc$.
The value of $\xc$ is determined by the point where locus of initial conditions ($\dot{x}+x^2=0$ in this case, represented by a green line in \Fig{fig:portrait}) meets the critical trajectory separating the two above-mentioned attractor basins.
This critical trajectory can in turn be found by integrating the ODE backwards in time, starting at a suitably chosen point in the vicinity of the attractor at the origin.

The purpose of this work is to develop approximate analytical solutions to particle trajectories obeying \Eq{eq:ode}, especially along the critical trajectory, and for the critical value $\xc$ relevant to filtration problems.  An exact solution of \Eq{eq:ode} presents a considerable challenge owing to the nonlinearity of the $x^2$ term, so we will employ asymptotic techniques to find the true solution perturbatively,
where renormalisation group (RG) approaches to solving singular ODEs turn out to be ideally suited to this problem \cite{chen1994, *chen1996, kirkinisPRE2008, kirkinisJMP2008, deville2008}; \textcite{kirkinisPRE2008} even illustrated the technique for our specific problem, deriving (but not solving) the RG equations.
Indeed, the simplicity of \Eq{eq:ode} makes it well-suited for a pedagogical introduction to asymptotic methods.

In section \ref{sec:perturbative} we set up the problem so that it can be solved perturbatively, and deduce the leading composite solution within a matched asymptotics approach.  The perturbative solution contains so-called ``secular terms'' which prevent refining the solution beyond leading order without more advanced techniques.  In section \ref{sec:rg} we deploy RG techniques to go beyond the composite solution.  Sections \ref{sec:perturbative} and \ref{sec:rg} proceed pedagogically for readers unfamiliar with asymptotics. In the remaining sections we use the results to deduce properties of the critical trajectory relevant to filtration dynamics.

\section{Perturbative approach}\label{sec:perturbative}
As we observed in the introduction, it would be deeply challenging to solve \Eq{eq:ode} in its raw form.  However, if we could temporarily turn off one of the terms then the remaining two terms become solvable.  The idea of a perturbative approach here is to start from a limit were we \emph{can} solve it (\eg\ dropping the nonlinear term), and then iteratively work out the perturbations required to construct the full solution.  To illustrate, we start from the assumption that $x$ is small by defining $x = \epsilon x_\Inner$ where $\epsilon$ is a small parameter.  The subscript $\Inner$ stands for ``inner'' which will be explained below.  With this substitution, the nonlinear term becomes a small correction:
\begin{equation}\label{eq:inner}
  \ddot{x}_\Inner + \dot{x}_\Inner = - \epsilon x_\Inner^2\,.
\end{equation}
The left-hand side now contains the leading \emph{linear} terms, which are solvable as $\epsilon \to 0$ with standard techniques.  The choice of terms leading terms is referred to as the ``dominant balance''; in this case $\ddot{x} + \dot{x} = 0$ is the dominant balance.

To construct a \naive\ perturbation series solution in $\epsilon$, we assume
\begin{subequations}\label{eq:naive-solution}
\begin{equation}\label{eq:naive-series}
  x_\Inner(t)
  =
  x_\Inner^{(0)}(t) + \epsilon x_\Inner^{(1)}(t) + \order{\epsilon^2}\,,
\end{equation}
where $x_\Inner^{(i)}(t)$ are \emph{required} to be $\order{1}$ functions because the fundamental assumption of an asymptotic series expansion is that terms appear in decreasing order of magnitude.  If one function became $x_\Inner^{(i)} = \order{\epsilon^n}$ for some $|n| \ge 1$ then this would invalidate the assumed ordering of terms by magnitude and thereby invalidate the series solution itself.  Inserting this series into \Eq{eq:inner} and solving at each order in $\epsilon$ gives the leading terms in the series:
\begin{align}
  \label{eq:leading-inner}
  x_\Inner^{(0)}
  &=
  A + B e^{-t}\,,
  \\
  x_\Inner^{(1)}
  &=
  -A^2 t + 2 A B e^{-t} (1 + t) - \frac{B^2}{2} e^{-2t}\,.
\end{align}
\end{subequations}
The first-order correction $x_\Inner^{(1)}$ contains a so-called `secular term' which limits the validity of the approach to short-times.  Specifically, the $-A^2 t$ term can become arbitrarily large at long-times.  In particular, $x_\Inner^{(1)}$ becomes $\order{\epsilon^{-1}}$ as $t \to 1/\epsilon$ which violates the assumptions of the series solution \Eq{eq:naive-series}.  As such, the solution \Eq{eq:naive-solution} is only valid at short-times for $t \sim o(\epsilon^{-1})$. 
When we say that this is the ``inner'' solution, we mean that it exists inside this temporal region (a ``boundary layer'') where the nonlinear term in \Eq{eq:inner} can be considered as a small perturbation.

Perhaps the most well-known method to go beyond the inner solution is by matched asymptotics.
In this approach we find a solution in a different limit, with a different dominant balance, and then construct a composite solution that contains both solutions. 
Let us look for a long-time solution by writing $x(t) = \epsilon x_\Outer(\tau)$ with $\tau = \epsilon t$.  The subscript $\Outer$ indicates these are solutions at longer times (the ``outer'' solution).  The particle equation of motion \Eq{eq:ode} becomes
\begin{equation}\label{eq:outer}
  \epsilon \partial_{\tau\tau}^2 x_\Outer + \partial_\tau x_\Outer + x_\Outer^2 = 0\,.
\end{equation}
Now the inertia term represents a small perturbation.  In the inertialess limit $\epsilon \to 0$, the solution to the dominant balance equation $\partial_\tau x_\Outer = -x_\Outer^2$ is
\begin{equation}\label{eq:leading-outer}
  x_\Outer = \frac{1}{\tau - C} + \order{\epsilon}\,,
\end{equation}
where $C$ is an integration constant.  This must match the leading inner solution \Eq{eq:naive-solution} (\ie\ dropping the $\order{\epsilon}$ term) at some intermediate time, \ie
\begin{equation*}
  \lim_{t \to \infty} \epsilon x_\Inner^{(0)} = \epsilon A = \lim_{\tau \to 0} x_\Outer = - \frac{1}{C}\,.
\end{equation*}
A composite solution can be constructed from the leading inner \Eq{eq:leading-inner} and outer \Eq{eq:leading-outer} solutions \via\ $x/\epsilon = x_\Inner^{(0)} + x_\Outer^{(0)} - A$, giving
\begin{equation}\label{eq:matched-asymptotics-solution}
  x(t) = \frac{\epsilon}{\epsilon t - C} + \epsilon B e^{-t} + \order{\epsilon^2}\,,
\end{equation}
which is uniformly valid across both time regimes, albeit with an $\order{\epsilon^2}$ error.

Matched asymptotics cannot be extended beyond leading order because it relies on being able to take the limit ${t \to \infty}$ limit of $x_I^{(0)}$ during matching of the solution in each regime.  The secular term in $x_I^{(1)}$ clearly prohibits this limit from converging to a finite number.  Supposing we want particle trajectories to some arbitrary precision (\eg\ to calculate $\xc$), we will therefore need more sophisticated techniques than matched asymptotics to go beyond leading order.

\section{Renormalisation group}\label{sec:rg}
The renormalisation group (RG) approach begins by reinterpreting the initial conditions as slowly varying amplitudes \cite{chen1994, *chen1996}.
This is achieved by incorporating higher-order terms with the same form as the leading solution.
In our case, we define
\begin{subequations}\label{eq:rg-rescalings}
  \begin{align}
    \widetilde{A}(t) &= A(1 - A \epsilon t + \order{\epsilon^2})\,, \\
    \widetilde{B}(t) &= B(1 + 2A(1 + t) \epsilon + \order{\epsilon^2})\,,
  \end{align}
\end{subequations}
so that
\begin{equation}
  x_\Inner = \widetilde{A}(t) + \widetilde{B}(t) e^{-t}
  - \epsilon \frac{\widetilde{B}(t)^2}{2} e^{-2t} + \order{\epsilon^2}\,,
\end{equation}
with the secular terms absorbed into the amplitudes.
The solution remains uniformly valid (even at long-times) when the slowly varying amplitudes $\widetilde{A}(t)$ and $\widetilde{B}(t)$ satisfy RG equations, which can be shown through a number of equivalent ways.

The method of the original papers \cite{chen1994, *chen1996} followed developments of field theoretic calculations (in particular by introducing the equivalent of an ``ultraviolet cutoff'').
Their method was conceptually subtle, but in practice becomes mechanistically simple to implement.
They would first redefine the \emph{constant} coefficients $A$ and $B$ such that $t \to t - t_0$ in the definitions of $\widetilde{A}(t) \to \widetilde{A}(t_0; t)$ and $\widetilde{B}(t) \to \widetilde{B}(t_0; t)$ in \Eq{eq:rg-rescalings}.
The RG equations follow from $\widetilde{A}'(t_0) = \widetilde{B}'(t_0) = 0$ so that the final solution does not depend on these auxiliary times.
In effect, this solution determines the form of amplitudes required to such that $t - t_0$ remains small even as $t \to \infty$.
We could then interpret the auxiliary variable $t_0$ as a slowly varying reference time which is ultimately absorbed into the amplitude by solving the RG equation.
\textcite{ziane2000} observed equivalences between this approach and the theory of averaging, resummation and Poincar\'e normal forms.
\textcite{kirkinisPRE2008} developed an alternative method to derive the RG equations without introducing the auxiliary variable $t_0$ that is conceptually simpler but involves more algebraic manipulations.

Whichever method is employed, secular terms are eliminated when the amplitudes satisfy the RG equations:
\begin{subequations}\label{eq:leading-rg}
  \begin{align}
    \frac{\widetilde{A}'}{\widetilde{A}} &= -\epsilon \widetilde{A} + \order{\epsilon^2}\,, \\
    \frac{\widetilde{B}'}{\widetilde{B}} &= 2\epsilon \widetilde{A} + \order{\epsilon^2}\,.
  \end{align}
\end{subequations}
These RG equations play the equivalent role of outer equations in matched asymptotics.
\textcite{kirkinisPRE2008} obtained the RG equations \Eq{eq:leading-rg} but did not solve them.
The solution correct to $\order{\epsilon}$ is
\begin{subequations}
  \begin{align}
    \widetilde{A} &= \frac{1}{\epsilon t - C} + \order{\epsilon^2}\,, \\[3pt]
    \widetilde{B} &= (\epsilon t - C)^2 D + \order{\epsilon^2}\,,
  \end{align}
\end{subequations}
with new constants of integration $C$ and $D$.  Inserting these back into \Eq{eq:naive-solution} and substituting $x = \epsilon x_\Inner$ gives the uniformly valid solution
\begin{equation}\label{eq:rg-solution}
  \begin{split}
    x(t) = \frac{\epsilon}{\epsilon t - C} &+ \epsilon D (\epsilon t - C)^2 e^{-t}\\
    &\qquad{} - \epsilon^2 \frac{D^2}{2} (\epsilon t - C)^4 e^{-2t} + \order{\epsilon^3}\,.
    \end{split}
\end{equation}
Note that this is identical to the composite solution from matched asymptotics when truncated at $\order{\epsilon}$, \ie
\begin{equation*}
  x(t) = \frac{\epsilon}{\epsilon t - C} + \epsilon D C^2 e^{-t} + \order{\epsilon^2}
\end{equation*}
which is the same as \Eq{eq:matched-asymptotics-solution} with $B = D C^2$.

Importantly, we have included the subleading term proportional to $e^{-2\tau}$ in \Eq{eq:rg-solution}; the original papers developing this route (by physicists) tend to ignore the subleading contributions (\eg\ \Refcite{chen1994, *chen1996}), whereas in mathematical literature these are normally included \cite{kirkinisPRE2008, *kirkinisJMP2008, *deville2008}.
We will see in later sections that it is is essential to include this term in order to obtain accurate solutions around the separatrix.

\section{Leading approximation of the separatrix}
\subsection{Obtaining the critical trajectory}
The previous two sections outline how to obtain solutions perturbatively \via\ two asymptotic methods.  Of these, the RG approach is capable of being systematically extendable beyond leading order, in principle to arbitrary precision.
However, our purpose in deriving these solutions is to obtain insight into the nonlinear separatrix which characterises this system, and these leading order solutions are sufficient to provide qualitative insight into the dynamical transition.

The long time behaviour of the leading-order solutions given in Eqs.\ \eqref{eq:matched-asymptotics-solution} and \eqref{eq:rg-solution} depends on the sign of $C$.
For $C > 0$ we see divergence $x \to -\infty$ as $x \to C$, implying a change of sign for some values of $D$.
If we redefine $B = D C^2 - \epsilon D^{2} C^{4} / 2$ in the RG solution \Eq{eq:rg-solution}, then for both solutions we find
\begin{equation}\label{eq:x0}
  \frac{x_0}{\epsilon} \equiv \frac{x(t=0)}{\epsilon} = -\frac{1}{C} + B\,,
\end{equation}
so valid initial conditions for particle trajectories $x_0 > 0$ require $B > {1}/{C}$.
When $C < 0$ then $x \sim {t}^{-1}$ in the long time limit, so we see decay to the origin for this choice of initial conditions.
For $C > 0$ the particle diverges to $x \to -\infty$ as $t \to C / \epsilon$, implying crossing $x = 0$ within finite time.
We therefore see collisions for $C \ge 0$.

The separatrix occurs at $C \to \pm \infty$, so $1/C$ seems to be the more natural variable to work with.
The separatrix occuring at $1/C = 0$ is defined at leading order by
\begin{equation*}
  x(t) \sim e^{-t} + \order{\epsilon^2}\,,
\end{equation*}
or $x(t) = -x'(t)$ approaching the origin \footnote{Incidentally, this exactly matches the result of a linear stability analysis about the origin which is perhaps unsurprising as the leading-order describes linear behaviour.}. The separatrix correct to $\order{\epsilon}$ is then a straight line about the origin. Extending this analysis away from the origin to derive the fully nonlinear form of the separatrix requires a more refined approach.

\subsection{Break-down of leading-order solution occurs because assumptions are physically inconsistent}
Before extending to higher-order, we can gain a little more insight by determining the constants in terms of the other initial condition in $x'(t=0)$.
We will see that a solution can only be found within a restricted region of phase space.Far from being solely a mathematical subtlety of asymptotics, the break-down of the approach is indicative of a failure in our \emph{physical} assumptions.

We focus on the leading-order matched asymptotic solution \Eq{eq:matched-asymptotics-solution} because it is cleaner.
The initial velocity is
\begin{equation*}
  \frac{u_0}{\epsilon} \equiv \frac{x'(t=0)}{\epsilon}
  =
  -\frac{\epsilon}{C^2} - B\,,
\end{equation*}
This has solution
\begin{equation}\label{eq:matched-asymptotics-C}
  \frac{C}{\epsilon} = -\frac{1 + \sqrt{1 - 4 x_0 - 4 u_0}}{2(x_0 + u_0)}\,.
\end{equation}
$B$ can be caclulated by inserting this $C$ into \Eq{eq:x0}, but the explicit expression is long and so we omit it.
There is no real solution when $1 - 4 x_0 - 4 u_0 < 0$.
This condition is met when
\begin{equation*}\label{eq:u0-failure}
  u_0 > \frac{1 - 4 x_0}{4}\,.
\end{equation*}
This sets limits on applicability of the approach.
To understand why, it is helpful to consider approaching the $\dot{x}$-axis where this condition reduces to $x_0 > 1/2$.
Along these set of initial conditions, $\dot{x}$ is small relative to $x^2$ which violates the dominant balance assumption of \Eq{eq:inner}.
The breakdown of the uniformly valid solution provides us with a clue about the relevant physics.
The correct dominant balance for these initial conditions would be $x^2 + \ddot{x} = 0$ which involves elliptic functions which are cumbersome as a starting point.

The take-away from the above is that the asymptotic solutions are not valid for all choices of initial conditions (at least to leading order) because they may develop an internal inconsistency.
Fortuitously, there is almost always a real solution with the zero initial acceleration condition, \ie\ when $u_0 = - x_0^2$ we have
\begin{equation*}
  \frac{C}{\epsilon} = -\frac{1 + \sqrt{( 1 - 2 x_0 )^2}}{2 x_0 (1 - x_0)}\,,
\end{equation*}
valid when $x_0 \ne 1$.
The term inside the square root is always greater than or equal to zero.
This initial condition is the standard one assumed within the filtration literature and so it is a reasonable one to consider.
Moreover, we hypothesise that this particular choice of initial condition is the most physical for porous media as large deviations from the flow field will likely be suppressed from capturing events.

\section{Separatrix beyond leading order}
\subsection{An exact recurrence relation}

\begin{figure}
  \centering
  \includegraphics[width=\linewidth]{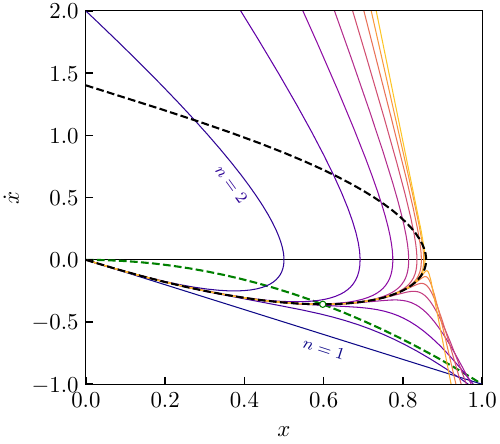}
  \caption{
    Approximation of the separatrix (dashed black line) via the asymptotic solution of \Eq{eq:separatrix-expansion} (solid lines, coloured purple $\to$ orange with increasing $n$).
    Contributions in \Eq{eq:separatrix-expansion} alternate in sign, so solutions tending upwards (downwards) are for even (odd) $n$.
    The approximations improve with $n$ in the region where $u < 0$; they break down as $u \to 0$ because this limit represents the breakdown of the approach where the dominant balance assumed in \Eq{eq:inner} is violated.
    The $\ddot{x} = 0$ nullcline (green dashed line) intersects the separatrix at the critical point $x = \xc$ in the region where the approximation seems to converge.
  }
  \label{fig:separatrix-asymptotic-solutions}
\end{figure}

Now we have enough insight to construct solutions beyond leading order.  We found that the separatrix occurs at $1/C = 0$, which zeros the amplitude $A \to 0$. Postulating that this remains true to higher order, we only look for solutions generated from the $e^{-t}$ leading solution.  In \Eq{eq:rg-rescalings} we found that
\begin{equation}
  \widetilde{B}(t) = B (1 + f(A))
\end{equation}
with $f(0) = 0$.  We find this remains true beyond leading order and so the amplitude equations are very simply $\widetilde{B} = B$ at arbitrary order when $A = 0$.  We have therefore conveniently lost the secular terms and can obtain better approximations to the separatrix through a \naive\ perturbation solution of \Eq{eq:inner}.
Going forward we will drop the $\Inner$ subscript from $x_\Inner$ to simplify notation, so we write the `inner' equation \Eq{eq:inner} as simply
\begin{equation}
  \ddot{x}+\dot{x}=-\epsilon x^2\,.\label{eq:naive}
\end{equation}
At the end we can set $\epsilon = 1$ to find an approximate solution for our original equation \Eq{eq:ode}.

Consider the asymptotic series
\begin{equation}
  x(t) = \sum_{n=0}^\infty (-\epsilon)^k x_n(t)
\end{equation}
where $x_n = \order{1}$.  Substituting this into \Eq{eq:naive} indicates that $x_n(t)$ should solve
\begin{equation}\label{eq:naive-odes}
  \ddot{x}_n + \dot{x}_n + \sum_{k=0}^{n-1} x_k x_{n-1-k} = 0\,.
\end{equation}
This presents a set of linear ODEs to be solved order-by-order for $x_n(t)$, similar to the way we solved \Eq{eq:inner} for the `inner' solution in the matched asymptotics method, except that we should drop the secular terms for the critical trajectory.  With this in mind we try solutions of the form
\begin{equation*}
  x_n = B_n e^{-(n+1) t}\,,
\end{equation*}
where $B_0 = B$ the leading order constant (which corresponds to an arbitrary rescaling of the time-origin so we can choose $B=1$ without loss of generality).  This leads to a recurrence relation for the coefficients which we can express as
\begin{equation}\label{eq:B-recurrence}
  B_0=1\,,\quad
  B_{n+1} = \frac{1}{(n+1)(n+2)}\sum_{k=0}^n B_k B_{n-k}
\end{equation}
where the second expression defines $B_n$ for $n\ge1$.  The first few of these coefficients starting from $n=0$ are
\begin{equation}
  \{B_n\} = 
  \{1,\,\smallfrac{1}{2},\,\smallfrac{1}{6},\,\smallfrac{7}{144},\,
  \smallfrac{19}{1440},\,\smallfrac{37}{10\,800},\,\smallfrac{29}{33\,600},\,
  \cdots\}\,.
\end{equation}
The full solution for the separatrix can then be written
\begin{equation}\label{eq:separatrix-expansion}
  x(t) = z \sum_{n=0}^\infty B_n (-\epsilon z)^n\,,\quad z\equiv e^{-t}\,.
\end{equation}
As we do not have a closed-form for $B_n$, this must be truncated at some finite maximum value of $n$ to obtain an asymptotic approximation.
In \Fig{fig:separatrix-asymptotic-solutions} we show the exact separatrix (dashed red) compared against the results of \Eq{eq:separatrix-expansion} with increasing truncations.  As more terms in the series are included, we obtain increasingly good agreement with the exact result suggesting that this is in fact a converging series (albeit very slowly).

Note that for $n>4$ the accumulation point representing the intersection of the separatrix with the nullcline is rapidly approached.  This suggests that we can use the power series in \Eq{eq:separatrix-expansion} to obtain a series solution for the location of the intersection point itself, eliminating the need to integrate any ODEs.  This proves to be correct (section \ref{subsec:ser} below).

\subsection{Relationship to other integer sequences}\label{sec:catalan}
Before we proceed with this, let us compare the recurrence relation for the $\{B_n\}$, \Eq{eq:B-recurrence}, with similar sequences.  This convolutional recurrence relation is similar to the recurrence relation for the famous Catalan numbers $\{C_n\}$:
\begin{equation}\label{eq:C-recurrence}
  C_0=1\,,\quad
  C_{n+1} = \sum_{k=0}^{n} C_k C_{n-k}\,,
\end{equation}
where the second expression defines $C_n$ for $n\ge1$.  Our relation differs only by the $1/n(n+1)$ scaling.  The generating function of the Catalan numbers \Eq{eq:C-recurrence} is
\begin{equation*}
  C(t) = \frac{1 - \sqrt{1 - 4t}}{2t}
\end{equation*}
giving
\begin{equation*}
  C_n = \frac{(2n)!}{(n+1)!\, n!} \in \mathbb{N}\,.
\end{equation*}
From this it is evident that the recurrence relation for $C_n$ can be expressed as a single sum, rather than the double sum appearing in \Eq{eq:C-recurrence}.
It is possible that a similar transformation can be achieved for our recurrence relation \Eq{eq:B-recurrence}, suggesting that there may be a closed form for $x(t)$.

The coefficients $\{B_n\}$ do not appear to be a previously known sequence.  To see this, we transform these rational numbers into a sequence of integers by defining $b_n = n!\, (n+1)!\, B_n$ with the recurrence
\begin{equation*}
  b_0=1\,,\quad b_{n+1}
  = \frac{1}{n} \sum_{k=0}^n {n \choose k} {n \choose n-k} b_k b_{n-k}
  \in \mathbb{N}.
\end{equation*}
This gives the integer sequence
\begin{equation*}
  \{b_n\} = \{1,\> 1,\> 2,\> 7,\> 38,\> 296,\> 3132,\> \cdots \}
\end{equation*}
which does not have a pre-existing entry in the Online Encylopedia of Integer Sequences (OEIS) \cite{OEIS} at time of writing. A sequence $\{d_n\}$ generated from the very similar recurrence
\begin{equation*}
  d_0=1\,,\quad d_{n+1}
  = \sum_{k=0}^n {n \choose k} {n \choose n-k} d_k d_{n-k}
  \in \mathbb{N}\,.
\end{equation*}
\ie\ without the $1/n$ prefactor, produces
\begin{equation*}
  \{d_n\} = \{1,\> 1,\> 4,\> 33,\> 456,\> 9460,\> 274\,800,\> \cdots \}
\end{equation*}
which is the known sequence A002190 in the OEIS \cite{OEIS}.  This sequence is generated from
\begin{equation*}
  \sum_{n=0}^\infty \frac{d_n}{(n!)^2} t^n
  = - \ln{J_0(2\sqrt{t})}
\end{equation*}
where $J_0$ is the zeroth-order Bessel function of the first kind.  The equivalent ordinary generating function for $D_n = d_n / (n! (n+1)!)$ would therefore be
\begin{equation*}
  \sum_{n=0}^\infty D_n t^n = -\frac{1}{t} \int_0^t \dd z \, \ln{J_0(2\sqrt{z})}\,.
\end{equation*}
It is possible that a similar generating function can be obtained for \Eq{eq:B-recurrence} in closed form but we have not thus far succeeded in finding one.

In searching the literature for related sequences we became aware of an intriguing paper \cite{barry2021} by Barry where the author solves several convolution recurrences using Riordan group methods \cite{shapiro2022}. While the specific techniques in that paper do not seem to apply to our \Eq{eq:B-recurrence}, it is possible that a solution could be found \via\ the Riordan group more broadly.

\begin{table}
  \centering
  \begin{tabular}{rclrclr}
    \hline
    $n$ &\phantom{\quad}&
    \multicolumn{1}{c}{$\epsilon\zc$} & \multicolumn{1}{c}{term}
    &\phantom{\quad}&
    \multicolumn{1}{c}{$\epsilon \xc$} & \multicolumn{1}{c}{term} \\
    \hline
     1 && 1.00000000 &  1.000e-00 && 1.00000000 &  1.000e-00 \\
     2 && 1.00000000 &  0.000e-00 && 0.50000000 & -5.000e-01 \\
     3 && 0.91666667 & -8.333e-02 && 0.58333333 &  8.333e-02 \\
     4 && 0.90277778 & -1.389e-02 && 0.60416667 &  2.083e-02 \\
     5 && 0.91458333 &  1.181e-02 && 0.60138889 & -2.778e-03 \\
    10 && 0.91742317 &  4.125e-04 && 0.59786408 & -8.479e-05 \\
    15 && 0.91745309 &  2.015e-05 && 0.59777988 & -5.033e-06 \\
    20 && 0.91745195 &  1.117e-06 && 0.59777679 & -3.378e-07 \\
    25 && 0.91745176 &  6.525e-08 && 0.59777667 & -2.357e-08 \\
    30 && 0.91745174 &  3.872e-09 && 0.59777667 & -1.664e-09 \\
    \hline
    \end{tabular}
  \caption{The series for $\epsilon\zc$ and $\epsilon \xc$ from \Eqs{eq:zcser} and~\eqref{eq:xcser} respectively, truncated after the indicated terms.}
  \label{tab:critical-x}
\end{table}

\subsection{Convergent series for critical inertia in filters}
\label{subsec:ser}
Now we return to the problem of using the asymptotic power series solution for the separatrix to determine the critical point $\xc$ where the separatrix intersects with the nullcline $u = -x^2$.  We approach this by substituting the power series into the nullcline, and matching terms.  First, differentiating \Eq{eq:separatrix-expansion} with respect to $t$ gives
\begin{equation}
  u = \dot{x} = - z \sum_{n=0}^\infty (n+1) B_n (-\epsilon z)^n\,.
\end{equation}
Second, squaring the power series gives
\begin{align}
  x^2
  &=
  z^2 \sum_{n=0}^\infty \sum_{m=0}^\infty B_n B_m (-\epsilon z)^{n+m}
  =
  z^2 \sum_{n=0}^\infty \sum_{m=0}^n B_m B_{n-m} (-\epsilon z)^n
  \nonumber \\ &{}\qquad =
  z^2 \sum_{n=0}^\infty (n+1)(n+2) B_{n+1} (-\epsilon z)^n\,,
\end{align}
where we rewrote the double-sum and exploited the recurrence relation definition of $B_n$ to reduce this to a single sum.  Substituting these into $u = -x^2$ now gives
\begin{equation}
  \begin{split}
    \zc \sum_{n=0}^\infty (n+1) &B_n (-\epsilon \zc)^n \\
    & = \zc^2 \sum_{n=0}^\infty (n+1)(n+2) B_{n+1} (-\epsilon \zc)^n\,,
\end{split}
\end{equation}
where $\zc$ is the value of $z$ at the intersection point.  After changing limits of the right-hand side this simplifies to 
\begin{equation}\label{eq:separatrix-intersection}
  \sum_{n=0}^\infty (n+1) B_n (-\epsilon \zc)^n =
  -\sum_{n=1}^\infty n(n+1) B_n (-\epsilon \zc)^n\,.
\end{equation}
Anticipating that $\epsilon$ should drop out from the solution \footnote{Either by noting that $\epsilon$ only appears on both sides as the product $\epsilon \zc$, so we can choose to consider their product as the relevant variable.
Alternatively, we could recall from the section \ref{sec:perturbative} that transforming $x \to x/\epsilon$ would eliminate $\epsilon$ from the evolution equation.}, we aim to solve this for the product $\epsilon \zc$.

The above equation is not immediately useful because $\epsilon \zc = \order{1}$ so we cannot construct a series solution.
To construct a series we introduce a prefactor $\theta$ on the left hand side of $u=-x^2$ as a bookkeeping device.
Our aim is to solve $\theta u = -x^2$, with the understanding that our final solution is achieved by setting $\theta=1$.  With this \Eq{eq:separatrix-intersection} becomes
\begin{equation}\label{eq:theta}
  \theta \sum_{n=0}^\infty (n+1) B_n (-\epsilon \zc)^n =
  -\sum_{n=1}^\infty n(n+1) B_n (-\epsilon \zc)^n\,.
\end{equation}
We expand the product $\epsilon\zc$ in powers of $\theta$,
\begin{equation}
  \epsilon \zc = \sum_{n=1}^\infty a_n \theta^n\,,
\end{equation}
noting that the leading order term is $\theta$.  The coefficients $a_n$ are to be determined by inserting this series into \Eq{eq:theta} and solving the resulting linear algebraic equations order-by-order.  This is most readily done with the aid of a computer algebra system (\eg\ Mathematica), giving (setting $\theta=1$)
\begin{equation}
  \label{eq:zcser} 
  \epsilon \zc = 1 + 0 - \smallfrac{1}{12} - \smallfrac{1}{72} + \smallfrac{17}{1440} + \smallfrac{119}{21\,600} - \smallfrac{949}{725\,760} + \cdots\,.
\end{equation}
Note $a_2 = 0$ because the asymptotic series solution for $x$ \Eq{eq:separatrix-expansion} truncated at $\order{\epsilon}$ does not intersect with the nullcline $u = -x^2$ (\cf\ green line in \Fig{fig:separatrix-asymptotic-solutions}).

Equation~\eqref{eq:zcser} is a series for the critical \emph{time} when the critical trajectory meets the nullcline.  To obtain the intersection point itself we substitute it into \Eq{eq:separatrix-expansion} to get a series for $\xc$.  The end result (again setting $\theta=1$) is
\begin{equation}
  \epsilon \xc = 1 - \smallfrac{1}{2} + \smallfrac{1}{12} + \smallfrac{1}{48} - \smallfrac{1}{360} - \smallfrac{17}{4320} - \smallfrac{43}{80\,640} + \cdots\,.\label{eq:xcser}
\end{equation}

\begin{figure}[t]
  \centering
  \includegraphics[width=\linewidth]{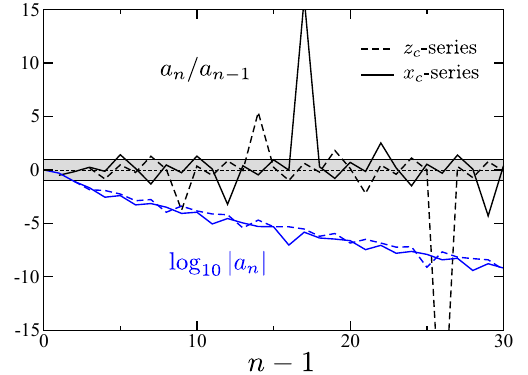}
  \caption{Asymptotic properties of the terms $a_n$ in \Eqs{eq:zcser} and~\eqref{eq:xcser} (using the same notation for both). The shaded band is $|a_n/a_{n-1}|<1$ and it is clear from the irregular excursions outside this that a simple d'Alembert ratio test for convergence is inconclusive.}
  \label{fig:asymptotics}
\end{figure}

Proceeding to higher orders, \Table{tab:critical-x} shows that the series in \Eq{eq:zcser} converges quite rapidly to $\epsilon\zc\simeq0.917452$, and likewise the series in \Eq{eq:xcser} converges to $\epsilon \xc\simeq0.597777$.  The latter numerically matches the value obtained from directly integrating \Eq{eq:ode} backwards in time from the origin.
Our more direct approach is more precise and accurate, as it does not depend on details of the quadrature method.
\Fig{fig:asymptotics} shows the convergence properties of the terms in both series.

\subsection{Asymptotic properties of the coefficients $B_n$}
Why do the series in \Eqs{eq:zcser} and~\eqref{eq:xcser} converge so relatively quickly?  We can gain some insight by considering the asymptotic properties of the coefficients $B_n$ in \Eq{eq:separatrix-expansion}.  We first notice that these should scale asymptotically as $B_n\sim n\times a^{-n}$ because if there are $n$ factors of $B_kB_{n-k}\sim k(n-k)a^{-n}$ in the sum in \Eq{eq:B-recurrence}, the numerator $\sim n^3a^{-n}$ whereas the denominator $(n+1)(n+2)\sim n^2$, thus preserving the scaling form for large $n$.  The constant $a$ can therefore be extracted in a Domb-Sykes plot \cite{hinch1991} by plotting $B_{n-1}/B_n$ as a function of $1/n$ and extrapolating to $1/n\to0$.

\begin{figure}[t]
  \centering
  \includegraphics[width=\linewidth]{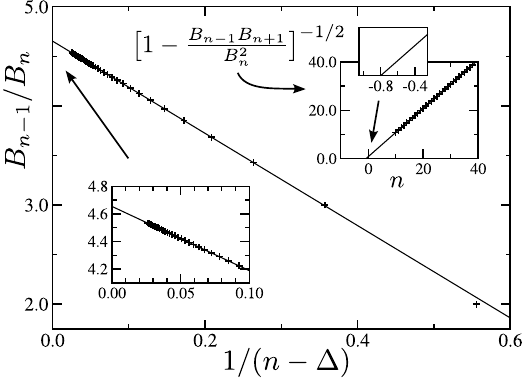}
  \caption{Domb-Sykes plots of $B_{n-1}/B_n$ versus $1/(n-\Delta)$ for the first 40 coefficients in the power series solution, \Eq{eq:separatrix-expansion}, for the critical trajectory.  The upper inset provides strong numerical evidence for the offset $\Delta=-4/5$ used to straighten out the curves in the main plot and lower inset.  The limiting value for $n\to\infty$ is $B_{n-1}/B_n\simeq4.6537$.\label{fig:domb-sykes}}
\end{figure}

We have found though that the results are improved by allowing for an offset $\Delta$ in the Domb-Sykes plot \cite{hinch1991}, or equivalently supposing that asymptotically $B_n\sim (n-\Delta) a^{-n}$.  The value of $\Delta$ can be found by plotting $[1-B_{n-1}B_{n+1}/B_n^2]^{-1/2}$ against $n$, which for large $n$ should asymptote to a straight line $\propto(n-\Delta)$.  This indeed works, as shown in the upper inset in \Fig{fig:domb-sykes}, and somewhat to our surprise the numerical evidence strongly indicates that $\Delta=-4/5$. Lacking a Ramanujan-like insight \cite{*[{E.g., the related Catalan numbers (section \ref{sec:catalan}) appear in many combinatoric problems, which was a particular area of expertise for Ramanujan; see }] [{}] hardy_1920}, the significance of this apparently exact result remains unknown to us at present.

The main Domb-Sykes plot shown in \Fig{fig:domb-sykes} is then made by plotting $B_{n-1}/B_n$ as a function of $1/(n-\Delta)$ with $\Delta=-4/5$, which yields $a\simeq 4.6537$.  The value $\epsilon\zc\simeq0.917$ (\Table{tab:critical-x}) where the critical trajectory meets the nullcline is considerably smaller than this, which means that at this point the oscillating terms $B_n(-\epsilon z)^n\sim n(-\epsilon z / a)^n$ in \Eq{eq:B-recurrence} shrink rather quickly since $\epsilon \zc / a\simeq 0.197$.  This explains the relatively rapid convergence of the series approximations to $\epsilon z_c$ and $\epsilon \xc$.

\section{Conclusion}
We have applied asymptotic techniques to perturbatively solve $\ddot{x} + \dot{x} + x^2 = 0$ close to the system's separatrix, which is the main object of interest in classifying dynamical behaviour for this system.  The series approximations for $\epsilon\zc$ and $\epsilon\xc$ given in \Eqs{eq:zcser} and~\eqref{eq:xcser} are the most striking results here, since they completely avoid numerical integration of the ODE and can be taken to arbitrary accuracy by iterating the defining relations.  This example is therefore perhaps something of a `poster child' for the application of RG methods to the analysis of nonlinear ODEs.  

While more realistic flow systems will feature additional complexity, this system constitutes the minimal toy system for understanding inertial particle collection onto a blunt obstacle close to its surface (where Reynolds number becomes small).
In particular, our analysis demonstrates provides a complete description of the critical inertia.
This can be seen through the transformed-time (``outer'') equation \Eq{eq:outer} where $\epsilon$ represents an effective inertia, commonly called the ``Stokes number''.
For a given initial starting position $x_0$, this critical inertia $\epsilon_c$ can be inferred from replacing $\epsilon \xc \to \epsilon_c x_0$ and the condition $\epsilon_c x_0 \simeq 0.597777$.

\begin{acknowledgments}
We would like to thank Richard Sear and Matthew Turner for stimulating conversations which inspired us to investigate this problem.
JFR acknowledges funding from the Alexander von Humboldt foundation.
\end{acknowledgments}

\bibliography{selected}

\end{document}